\documentclass[conference]{IEEEtran}
\usepackage{amssymb}
\usepackage{threeparttable}
\usepackage[cmex10]{amsmath}
\usepackage{caption}
\usepackage{subfigure}
\usepackage{graphicx}
\usepackage{graphics}
\usepackage{cite}

\usepackage[linesnumbered,ruled]{algorithm2e}
\usepackage{booktabs}
\usepackage{amsfonts,amssymb}
\usepackage{multicol,balance}
\usepackage{times}
\usepackage{stfloats}
\usepackage{setspace}
\usepackage{caption}
\usepackage{amsmath}
\usepackage{mathtools}

\usepackage[caption=false]{subfig}

\usepackage{url}

\makeatletter
\def\endthebibliography{%
  \def\@noitemerr{\@latex@warning{Empty `thebibliography' environment}}%
  \endlist
}
\makeatother

\begin{document}

\title{Privacy Preservation in Location-Based Services: A Novel Metric and Attack Model}
\IEEEoverridecommandlockouts
\author{
    \IEEEauthorblockN{Sina Shaham, Ming Ding, Bo Liu, Zihuai Lin, Jun Li}\\
    \IEEEauthorblockA{
        School of Electrical and Information Engineering, The University of Sydney, Australia \\
        Department of Engineering, La Trobe University, Australia\\
        Email: \{sina.shaham, zihuai.lin\}@sydney.edu.au, ming.ding@data61.csiro.au, b.liu2@latrobe.edu.au, jun.li@njust.edu.cn}
}

\maketitle

\begin{abstract}
Recent years have seen rising needs for location-based services in our everyday life. Aside from the many advantages provided by these services, they have caused serious concerns regarding the location privacy of users. An adversary such as an untrusted location-based server can monitor the queried locations by a user to infer critical information such as the user's home address, health conditions, shopping habits, etc. To address this issue, dummy-based algorithms have been developed to increase the anonymity of users, and thus, protecting their privacy. Unfortunately, the existing algorithms only consider a limited amount of side information known by an adversary which may face more serious challenges in practice. In this paper, we incorporate a new type of side information based on consecutive location changes of users and propose a new metric called transition-entropy to investigate the location privacy preservation, followed by two algorithms to improve the transition-entropy for a given dummy generation algorithm. Then, we develop an attack model based on the Viterbi algorithm which can significantly threaten the location privacy of the users. Next, in order to protect the users from Viterbi attack, we propose an algorithm called robust dummy generation (RDG) which can resist against the Viterbi attack while maintaining a high performance in terms of the privacy metrics introduced in the paper. All the algorithms are applied and analyzed on a real-life dataset.
\end{abstract}

\section{Introduction}
\label{section: 1}

With the ubiquitous use of smartphones and social networks, location-based services (LBSs) have become an essential part of the contemporary society. The users of smart devices can simply download location-based applications and query the information from the LBS provider. For example, LBSs offered by companies like Alibaba, Apple, and Google can be used to find nearby restaurants, track the parcels, and provide personalized weather notifications. The annual market for LBSs is expected to reach USD $77.84$ Billion by $2021$, with an annual growth rate of $38.9\%$ \cite{market}.

In spite of countless advantages of LBSs, the privacy issues associated with the user locations have raised many concerns in our society. An untrusted server can collect the location data of users and analyze it to learn sensitive information such as the type of queries submitted, shopping habits of users, and the address of users' properties or workplaces. Such information can be easily abused by the server or disclosed to other parties. Therefore, it is of great importance to devise new ways to preserve the location privacy of users defined as "the ability to prevent other parties from learning one's current or past locations" \cite{r1}.

The techniques to address the threats to the location privacy of users have attracted much attention among researchers \cite{n1,n2,n3,n4,r1}. Most of the literature is based on an approach called $k$-anonymity \cite{r2}. Using this criterion, the release of a location is said to provide $k$-anonymity, if the real location of any user is not distinguishable from at least $k-1$ other locations. Initially, the approach to hide the location of the user was conducted using a trusted anonymization server \cite{n5}, but later on, due to the shortcomings of this approach such as the anonymizer becoming the bottleneck itself, the use of dummy locations to achieve the $k$-anonymity was proposed in \cite{baseline}. Since then, the researchers have strived to develop dummy generation algorithms to preserve the $k$-anonymity for users.

The principal idea behind the dummy generation algorithms is to generate $k-1$ dummy locations aside from the real location of the user and submitting them all together to the LBS server while asking for a query from the LBS provider. Thus, it makes it difficult for an untrusted LBS provider, or so-called the adversary, to identify the real location of the user. The groundwork in this field was laid by the authors in \cite{baseline}. They generated the dummies randomly throughout the map and evolved them as users move. Followed by this work, the authors in \cite{grid1} and \cite{gridt} proposed to choose the candidate dummies from a virtual circle or grid constructed around the current location of the user. Unfortunately, in all of the mentioned works, the fact that the adversary might have some side information which can rule out the dummies or reveal the real location of the user was overlooked.

One important piece of side information which can be exploited by the adversary is the query probability of the locations across the map. The adversary can utilize the recorded data and infer the number of times that the users have queried over various locations on the map. Using this information, the adversary can calculate the query probability of each location, and then, identify the dummy locations according to the history of interests in locations. For instance, if a dummy has been chosen on a lake, where the query probability is basically close to zero, the adversary will then know with a high likelihood that such queried location is a dummy. And therefore, such naive selection of dummy locations compromises the location privacy of the user. To solve this issue, an enhanced algorithm was proposed by \cite{entropy}, referred to as the dummy-location selection (DLS) algorithm. Basically speaking, the authors used an entropy metric \cite{r3} to evaluate the queries submitted in different locations and generated the dummies in a way to maximize the entropy.

\begin{figure}[h]
\centering
\includegraphics[scale=0.5]{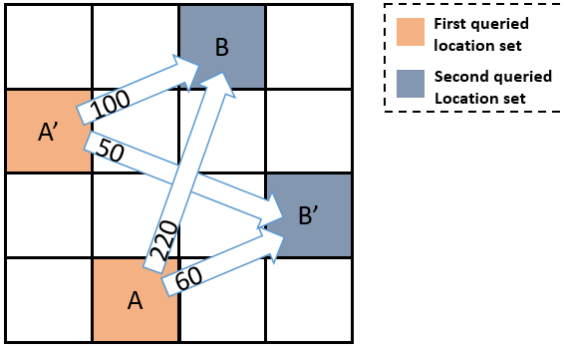}
\hspace{1em}
\centering
\caption{An example of location privacy of the user being compromised by considering the introduced side information.}
\label{f1}
\end{figure}

Although the DLS algorithm is promising for a stationary set of the queried locations including the real location and its associated dummies, the algorithm fails to address the privacy issues caused by the consecutive queries made to the LBS provider. In more detail, the authors have limited the side information to queries submitted in different locations but overlooked the fact the adversary has also access to the trajectories, and consequently, the number of times the paths between locations have been traveled. Having access to such extra side information, the adversary can expose the dummies and compromise the $k$-anonymity of the users. For further explanation, a toy example has been provided in Fig. \ref{f1},
where we show a user moving from location $A$ to location $B$ with $k$ set to two. The associated dummies of the real locations $A$ and $B$ are denoted by $A'$ and $B'$, respectively. The dummies in each location set are generated using the DLS algorithm, hence, they have a similar probability of being selected. The numbers on the directed edges indicate the number of times that users have queried the end location of the edge right after asking about the starting point of the edge. For instance, the users have queried location $B$ for $100$ times immediately after location $A'$. According to the DLS algorithm, the $k$-anonymity requirement has been preserved for each location. However, let us look at the four paths connecting the two sets of locations together and consider the number of times that each path has been inquired. It can be seen from Fig. \ref{f1} that location $B$ has been inquired for $320$ times after locations $A$ and $A'$ whereas location $B'$ has only received $110$ times of inquiries. Therefore, the adversary can infer with a high likelihood that the real location is possibly location $B$, and thus, compromise the location privacy of the user.

The main contributions of this paper follow.

\begin{itemize}
  \item We quantify the currently existing metric and name it cell-entropy and propose a new metric called transition for two consecutive queries which considers the introduced side information based on transitions of the users.
  \item We expand the transition-entropy metric for trajectories followed by developing two algorithms which can be applied on any of the existing dummy generation methods to improve the transition entropy.
  \item We propose an attack model based on the Viterbi algorithm and develop an algorithm to improve the resilience against the attack while maintaining the high performance in terms of cell-entropy and transition-entropy
  \item We analyze the performance of the proposed metrics and algorithms on a real-life dataset.
\end{itemize}

The rest of the paper is organized as follows. We start by explaining the existing works in literature in Section \ref{section: r}. Section \ref{section: 2} describes the system model used throughout the paper including the system architecture, the adversary model, and the side information used by the adversary. In section \ref{section:MetricS}, we introduce our proposed metrics followed by explaining the proposed attack model in section \ref{section: v}. Next, the proposed algorithms are illustrated in section \ref{section:AlgoS}. Finally, the analysis of the proposed metrics and algorithms is provided in section \ref{section:Performance}, and we conclude our work in section \ref{section: Conclusions}.


\section{Related works}
\label{section: r}
\textit{Anonymity} is defined as "the state of being not identifiable within a set of subjects, the anonymity set" \cite{s1}. Also, the location of a user is said to be $k$-anonymous if it is not distinguishable from at $k-1$ other user locations \cite{s2}. To obtain $k$-anonymity for users several approaches have been proposed in which we have identified four broad categories: location cloaking, mix-zones, pseudonyms, and dummy aided algorithms.

The research on location cloaking was initiated by Gruteser and Grunwald \cite{s3}. The key idea is to employ a trusted server to aid the users preserve their $k$-anonymity. Upon receiving a query from a user, the location anonymizer server computes a cloaking box including the location of the user and $k-1$ other user locations and queries the requested service from the LBS provider for all the $k$ locations. Therefore, making it difficult for the LBS provider to identify the user \cite{s4,s5}. Several algorithms have been proposed to implement location cloaking scheme such as ICliqueCloak \cite{s7} and MaxAccuCloak \cite{s6}. The main drawback of the location cloaking is the need for a location anonymizer which is an additional cost overhead to the system. The location anonymizer can become a bottleneck itself both from the privacy and computational complexity perspective.

The authors in \cite{s8} proposed the idea of mixed zones. Mixed zone is defined as the spatial zone where the identity of users is not identifiable. All the users entering into a mixed zone will change their pseudonym to a new unused pseudonym making it difficult for the adversary to identify the users. The anonymization process is performed by a middle-ware mechanism before transferring the data to third-party applications. The authors further extended their work in \cite{s9} by considering irregular shapes for mix zones. Moreover, the use of mix zones has particularly attracted attention in vehicular communications. Applying the mix zones method for road networks is considered in \cite{s10,ss10}, where a mixed zone construction method called MobiMix is proposed. Lu et al. \cite{w1} exploited the pseudonym changes to for mix- zones at social spots and Gao et al. \cite{w2} applied mix zones approach on trajectories for mobile crowd sensing applications. Furthermore, the use of cryptography for generation of mix zones in vehicular communications is considered in \cite{s11}. As it is the case for location cloaking approach, the main drawback of mix zones is also the need for a middle-ware mechanism or a trusted party before transferring the data to an untrusted LBS provider.

Another technique to increase the location privacy of the users is based on the assignment of pseudonyms to hide the identity of the users. The identity of a user can be the name of the person, a unique identifier such as IP address, or any properties that can be related to the user. The authors in \cite{g1} proposed a scenario so-called intermediary scenario in which a trusted intermediary collects the location information of the users such as GPS data and assigns a pseudonym before sending them to a third party LBS provider which is considered to be untrusted. The paper claims that the use of pseudonyms prevents the third party LBS provider from identifying and tracking the users. The work in \cite{g2} suggests that instead of delegating the generation of pseudonyms to the location intermediary, users generate the pseudonyms themselves. The use of pseudonyms for preserving the location privacy has also been considered in vehicular communication systems such as the work in \cite{g3}. There are several drawbacks associated with this approach. First of all, many of the location-based applications require the users to subscribe in order to use the offered services. Secondly, similar to the last two categories, this approach also requires a trusted intermediary. And finally, analyzing the patterns in location data an adversary can compromise the identity of the users \cite{g4}.

The last category which is considered to be a more promising approach since there is no need for a trusted anonymizer as it was the case for the location cloaking, mix zones, and pseudonyms is the use of dummy locations. This approach was initially proposed in \cite{baseline}. The principal idea is to achieve $k$-anonymity by sending $k-1$ dummy location aside from the real location of the user while asking for an LBS from an untrusted LBS provider. All the locations use the same identifier corresponding to the user. Having dummy locations, it would become difficult for the adversary to identify the real location of the users. Several algorithms have been proposed to help the users generate the dummies. The authors in \cite{grid1} proposed to use a virtual circle or a virtual grid which is based on the real location of users to generate the dummies. The idea was further developed in \cite{gridt}. More recently, an algorithm called dummy-location selection (DLS) was proposed in \cite{entropy}. The algorithm takes the number of queries made in the map into consideration and proves via simulations that previous algorithms are susceptible if the adversary exploits this side information. Although the algorithm provides a great framework for the generation of dummies, it does not take into account that the users are in danger of losing their location privacy if the adversary tracks them and access to other side information such as the number of transitions made in the map. Do et al. \cite{w3} utilized conditional probabilities to generate realistic false locations and Hara et al. \cite{w4} proposed a method based on physical constraints of the real environment.

\section{System Model}
\label{section: 2}

\subsection{System Architecture}

\begin{figure}[h]
\centering
\includegraphics[scale=.53]{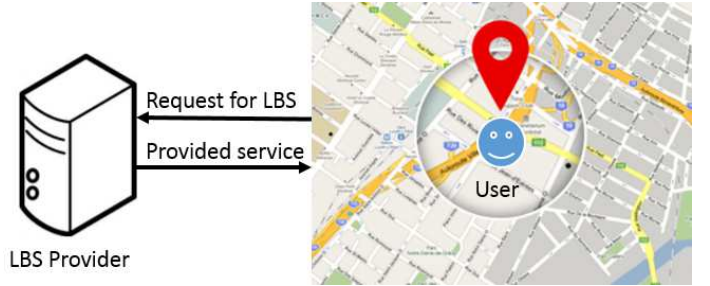}
\hspace{1em}
\centering
\caption{Non-cooperative system architecture for LBSs.}
\label{f2}
\end{figure}

In this paper, we adopt a non-cooperative system architecture \cite{r4}, as shown in Fig. \ref{f2}. In this architecture, the LBS users are directly in contact with the LBS provider with no middle-man or a third party service provider.

Assume that the location map is divided into an $n\times n$ grid and a user communicates with an LBS server for a service. At time $t^q$, the user intends to make his/her $q$-th query from the service provider, preserving $k^q$-anonymity. Here, $k^q$ quantifies the privacy protection requirement of the user. This metric implies that the adversary is not able to identify the real location of the user with a probability higher than $1/k^q$. Hence, such user needs to transmit $k^q-1$ dummy location to hide its true location from the observer. Note that by the term location we refer to the cell in which the user is located. We denote the set of locations transmitted to the LBS provider at $q$-th query by
\begin{equation}\label{e1}
LS^q=\{l^q_1,l^q_2,...,l^q_{k^q}\}.
\end{equation}
Also, the real location is shown by $r^q$ where $r^q \in LS^q$. The probability of location $l^q_x$ being the real location is shown by
\begin{equation}\label{e2}
\textrm{Pr}(l^q_x=r^q) \textrm{   for } x=1,...,k^q.
\end{equation}
In the next query, the user requires $k^{q+1}$-anonymity and queries the location set $LS^{q+1}=\{l^{q+1}_1,l^{q+1}_2,...,l^{q+1}_{k^{q+1}}\}$ from the LBS provider. The probability of $l^{q+1}_y \in LS^{q+1}$ being queried consecutively after $l^{q}_x \in LS^{q}$ is denoted by
\begin{equation}\label{e3}
\textrm{Pr}(l^{q}_x\Rightarrow l^{q+1}_y).
\end{equation}

\subsection{Adversary Model}

Two types of adversary models are considered in our work: an active adversary, and a passive adversary. The passive adversary can listen to the communication between the users and the LBS provider. Analyzing the collected information, the passive adversary can compromise the location privacy of the users by performing an eavesdropping attack. An active adversary, on the other hand, compromises the LBS provider and has access to the information stored on the server. In our work, the active adversary is assumed to be the LBS provider itself.

\subsection{Side Infromation}

The adversary is assumed to possess the location map of the area where the users are distributed. He has access to the queries made by the users and can record them over time to obtain the history of the locations where the users have queried from. Moreover, the adversary can calculate the query probability of different locations in the map, which is defined as the number of times a particular location has been queried. The adversary can exploit the query probability to infer the probability of a location being genuine or fake in the future queries. For instance, if a user queries two location, one with a comparably higher probability, it is more likely that the real location has the higher probability.

Apart from the possession of traditional side information by the adversary, we assume that the adversary has access to the number of times each path has been traveled on the map. The authorities do not provide any time limit for storing the location information of the users, as it is the case in the US \cite{act}. This lack of legislation enables the adversary to monitor the users and access to the trajectories traveled by them. Therefore, the adversary not only has the data on the number of queries made on each location, but it is well-aware of the number of times that a location has been queried consecutively after the other locations.


\section{Performance Metrics of Privacy}
\label{section:MetricS}
In this section, we briefly explain a metric which was partially developed in \cite{entropy}. Then, we propose a metric called transition-entropy to analyze the privacy preservation in LBSs for two consecutive queries followed by expanding the metric for trajectories.

\subsection{Cell-entropy Metric}

Although not mentioned as a metric in \cite{entropy}, cell-entropy was implicitly proposed as part of the DLS algorithm. We have named this metric cell-entropy to distinguish it from the transition-entropy metric proposed in this paper.
For a given location set $LS^q=\{l^q_1,l^q_2,...,l^q_{k^q}\}$ which includes the real location of a user and $k^q-1$ dummies chosen to preserve $k^q$-anonymity, the set of query probabilities are shown by $B^q=\{b^q_1,b^q_2,...,b^q_{k^q}\}$ where $b^q_j$ is the query probability of location (cell) $l^q_j$ for $j=1...k^q$. The query of probability of cell $l^q_j$ is calculated by

\begin{equation}\label{e4}
b^q_j=\dfrac{\textrm{number of queries in }l^q_j}{\textrm{number of queries in whole map}}.
\end{equation}
The cell-entropy borrows the concept of entropy from information theory to quantify the uncertainty in query probability of the locations in $LS^q$. The cell-entropy metric for location set $LS^q$ can be defined as \cite{entropy}
\begin{equation} \label{e5}
\centering
h= -\sum_{j=1}^{k^{q}} b^q_j \log_{2}b^q_j.
\end{equation}

\subsection{Transition-entropy}
The main purpose of the metric we propose here is to provide a benchmark for the comparison between dummy-based algorithms taking into account the comprehensive side information we consider in this paper. The metric indicates the susceptibility of the existing algorithms to attacks on location privacy of the users as the $k$-anonymity requirement of the users can easily be compromised in trajectories. Hence, necessitating the need for the development of new algorithms for preserving the location privacy of the users. We start by illustrating the metric for two consecutive queries and then generalizing it for trajectories.

\subsubsection{Transition-entropy metric for two consecutive queries} Assume that at time $t^q$ a user makes its $q$-th query and has an anonymity constraint of $k^q$, and requests the service for the location set of $LS^q=\{l^q_1,l^q_2,...,l^q_{k^q}\}$. The set $LS^q$ includes $k^q-1$ dummies and the real location of user. Then, at time $t^{q+1}$ the user moves to a new location with the anonymity constraint of $k^{q+1}$ and makes his $(q+1)$-th query providing the server with the location set of $LS^{q+1}=\{l^{q+1}_1,l^{q+1}_2,...,l^{q+1}_{k^{q+1}}\}$ consisting of the real location of the user and the associated dummies. The dummies can be generated using any of the existing algorithms.

\begin{figure}[h]
\centering
\includegraphics[scale=.61]{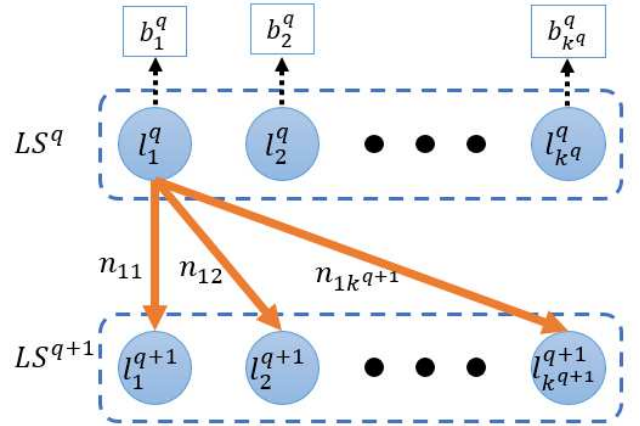}
\hspace{1em}
\centering
\caption{The bipartite graph generated by the consecutive queries of a user.}
\label{f3}
\end{figure}

Using the sets $LS^q$ and $LS^{q+1}$, we generate a bipartite graph shown in Fig. \ref{f3}, where each set forms the vertices at a side of the graph.
We denote the number of times the location $l^{q+1}_y \in LS^{q+1}$ follows the location $l^{q}_x \in LS^{q}$ by $n_{xy}$ and assign it to the directed edge connecting $l^{q}_x$ to $l^{q+1}_y$. Also, for every location $l^{q}_x \in LS^{q}$, we denote query probability of the location $l^{q}_x$ by $b^{q}_x$. The query probability of a cell is calculated by dividing the number of times that cell has been called over the whole number of queries of the map. This data is calculated from the history of data LBS provider holds.

We would like to find out how probable it is for each member of the location set $LS^{q+1}$ to be the real location of the user ($r^{q+1}$) given the location set $LS^{q}$ in the previous query from the LBS provider. In other words, the aim is to calculate the posterior probability of the members in $LS^{q+1}$ with respect to $LS^{q}$. This probability for each member of $LS^{q+1}$ can be calculated based on the $LS^{q}$ as
\begin{align} \label{e6}
\forall\  l^{q+1}_y \in LS^{q+1}&:\nonumber \\ &\textrm{Pr}(l^{q+1}_y=r^{q+1}|LS^q) =\\  &\sum_{s=1}^{k^{q}}\textrm{Pr}((l_{s}^{q}\Rightarrow l^{q+1}_y), (l^{q}_{s}=r^{q}))=\label{e7} \\
&\sum_{s=1}^{k^{q}}\textrm{Pr}(l_{s}^{q}\Rightarrow l^{q+1}_y|l_{s}^{q}=r^q) \textrm{Pr}(l^{q}_{s}=r^{q}),\label{e8}
\end{align}
where the equation (\ref{e7}) is the joint probability of $l^{q}_{s}$ being the real location of $LS^q$ and moving to the location $l^{q+1}_y$ after $l^{q}_{s}$. The former probability in equation (\ref{e8}) can be calculated as
\begin{align}
\forall\  l^{q+1}_y \in &LS^{q+1},\   \forall\  l^{q}_x \in LS^{q}:\nonumber  \\
&\textrm{Pr}(l_{x}^{q}\Rightarrow l^{q+1}_y|l_{x}^{q}=r^q)=\dfrac{n_{xy}}{\sum_{y=1}^{k^{q+1}} n_{xy}},\label{e9}
\end{align}
and the latter probability which indicates the normalized query probability as
\begin{align}
\forall\  l^{q}_x \in LS^{q}: \  \textrm{Pr}(l^{q}_{x}=r^{q})=\dfrac{b^{q}_x}{\sum_{j=1}^{k^q} b^{q}_j}.\label{e10}
\end{align}
Note that equation (\ref{e10}) indicates that the posterior probability of the cells in $LS^q$ are set to the normalized query probability of the locations. Calculating equation (\ref{e8}) for every member of the location set $LS^{q+1}$, the posterior probabilities of the locations in $LS^{q+1}$ are derived based on the $LS^q$. Having these probabilities, we exploit the concept of entropy to infer the uncertainty in identifying the dummies or the real location of the users calculated by
\begin{equation} \label{e11}
\centering
h= -\sum_{y=1}^{k^{q+1}} \textrm{Pr}(l^{q+1}_y=r^{q+1}|LS^q) \log_{2}\textrm{Pr}(l^{q+1}_y=r^{q+1}|LS^q).
\end{equation}
\begin{algorithm}[t]
\DontPrintSemicolon 
\textbf{Input:} The location sets $LS^{q}$ and $LS^{q+1}$.\\
\textbf{Output:} The transition-entropy of $LS^{q+1}$ with respect to $LS^{q}$.\\
\textbf{Initialization:}  $CellSum=0$, $h=0$.\\

\For {$1\leq x \leq k^{q}$} {
$EdgeSum=0$\\
\For {$1\leq y \leq k^{q+1}$} {
$EdgeSum=EdgeSum+n_{xy}$
}
\For {$1\leq y \leq k^{q+1}$} {
$\textrm{Pr}(l_{x}^{q}\Rightarrow l^{q+1}_y|l_{x}^{q}=r^q)=n_{xy}/EdgeSum$
}

}

\For {$1\leq x \leq k^{q}$} {
$CellSum= CellSum +b^{q}_x$
}
\For {$1\leq x \leq k^{q}$} {
$\textrm{Pr}(l^{q}_{x}=r^{q})=b^{q}_x/CellSum$
}
\For {$1\leq y \leq k^{q+1}$} {
$\textrm{Pr}(l^{q+1}_y=r^{q+1}|LS^q) =0$\\
\For {$1\leq x \leq k^{q}$} {
$\textrm{Pr}(l^{q+1}_y=r^{q+1}|LS^q)=\textrm{Pr}(l^{q+1}_y=r^{q+1}|LS^q)$\\
$\  \   \   \   \    \   \  \   \   \   +\textrm{Pr}(l^{q+1}_y=r^{q+1}|l_{x}^{q}=r^q) \textrm{Pr}(l^{q}_{x}=r^{q})$
}

$h= h -$\\
 $\   \   \    \textrm{Pr}(l^{q+1}_y=r^{q+1}|LS^q) \log_{2}(\textrm{Pr}(l^{q+1}_y=r^{q+1}|LS^q))$
}
\textbf{return} $h$
\caption{Calculation of transition-entropy for the location set $LS^{q+1}$ with respect to $LS^{q}$.}
\label{Algo1}
\end{algorithm}

\begin{algorithm}[t]
\DontPrintSemicolon 
\textbf{Input:} The location sets $LS^q,LS^{q+1}...,LS^{q+c}$.\\
\textbf{Output:} The TransitionEntropy of $LS^{q+c}$ with respect to $LS^q,LS^{q+1}...,LS^{q+c-1}$.\\
\textbf{Initialization:} TransitionEntropy $=0$.\\

Run Algo. 1 for $LS^q$ and $LS^{q+1}$\\
\For {$q+1\leq query \leq q+c-1$} {
Normalize posterior probabilities of $LS^{query}$\\
Query probabilities of $LS^{query}$ $\leftarrow$ posterior probabilities of $LS^{query}$\\
Run Algo. 1 for $LS^{query}$ and $LS^{query+1}$\\
}
Normalize posterior probabilities of $LS^{q+c}$\\
TransitionEntropy $\leftarrow$ calculate Entropy\\
\textbf{return} TransitionEntropy
\caption{Calculation of transition-entropy for trajectories of length $c+1$.}
\label{Algo2}
\end{algorithm}

We call $h$, the transition-entropy of the location set $LS^{q+1}$ with respect to $LS^{q}$. The transition-entropy metric represents the uncertainty of identifying the real location by the adversary in consecutive queries from the LBS provider. Having a larger value for the transition-entropy indicates that for each member of $LS^{q+1}$, the probability of the paths originating from the $LS^{q}$ to the destination of that member is similar to the other members of $LS^{q+1}$. Hence, it would be more difficult for the adversary to compromise the $k^{q+1}$-anonymity of the users based on the transitions made from their previous query. The formal algorithm for calculating the transition-entropy of the location set $LS^{q+1}$ with respect to $LS^{q}$ is presented in algorithm \ref{Algo1}.
The main advantages of the metric can be mentioned as: (i) considering the performance of the dummy-based algorithms in trajectories and not just a stationary set of locations; (ii) being able to investigate the performance of the dummy-based algorithms for users with varying $k$-anonymity requirements in their trajectory; (iii) entailing many other factors such as time reachability or direction similarity considered in other works.

\subsubsection{Transition-entropy metric for trajectories}

In this subsection, we generalize the transition metric for trajectories with different lengths. Assume that at time $t^{q+c}$ the user makes its ($c+1$)-th query providing the LBS provider with the location set $LS^{q+c}=\{l^{q+c}_1,l^{q+c}_2,...,l^{q+c}_{k^{q+c}}\}$ with privacy requirement of $k^{q+c}$. The previous queried location sets are shown by $k^{q+i}$ for $i=0,...,c-1$ each with the privacy requirement of $k^{q+i}$ and being queried at time $t^{q+i}$. Initially, our aim is to calculate the posterior probability of each location in $LS^{q+c}$. The posterior probabilities indicate the likelihood of any location in $LS^{q+c}$ being the real location of the user based on the previous queries that the user has made. Posterior probability for each location in $LS^{q+c}$ can be written as

\begin{align}
\forall\  &l^{q+c}_y \in LS^{q+c}:
\textrm{Pr}(l^{q+c}_y=r^{q+c}|LS^q,...,LS^{q+c-1}) =\label{ee13}  \\
&\quad  \sum_{s^{c}=1}^{k^{q+c-1}}\textrm{Pr}((l_{s^{c}}^{q+c-1}\Rightarrow l^{q+c}_y), \nonumber \\ &\quad \quad \quad \quad (l^{q+c-1}_{s^{c}}=r^{q+c-1})|LS^q,...,LS^{q+c-2}))=\\
&\quad  \sum_{s^{c}=1}^{k^{q+c-1}}\textrm{Pr}(l_{s^{c}}^{q+c-1}\Rightarrow l^{q+c}_y|l^{q+c-1}_{s^{c}}=r^{q+c-1})\times \nonumber \\
&\quad \quad \quad \quad  \textrm{Pr}(l^{q+c-1}_{s^{c}}=r^{q+c-1}| LS^q,...,LS^{q+c-2} ).\label{e13}
\end{align}
\begin{figure*}[!t] 
\normalsize
\begin{equation}\label{e12}
\mathlarger{\sum_{s^{c}=1}^{k^{q+c-1}}\sum_{s^{c-1}=1}^{k^{q+c-2}}...\sum_{s^{1}=1}^{k^{q}}(} \textrm{Pr}(l^{q}_{s^{1}}=r^{q})
\textrm{Pr}(l_{s^{c}}^{q+c-1}\Rightarrow l^{q+c}_y|l^{q+c-1}_{s^{c}}=r^{q+c-1})
\mathlarger{\prod_{i=1}^{c-1}} \textrm{Pr}(l_{s^{i}}^{q+i-1}\Rightarrow l^{q+i+1}_{q+i}|l^{q+i-1}_{s^{i}}=r^{q+i-1})
\mathlarger{)}
\end{equation}
\hrulefill
\end{figure*}
Following the same process of moving from equation~(\ref{ee13}) to equation~(\ref{e13}), the probability of $\textrm{Pr}(l^{q+c-1}_{s^{c-1}}=r^{q+c-1}| LS^q,...,LS^{q+c-2} )$ can be solved recursively to reach the equation (\ref{e12}) where the transition probabilities can be calculated similar to the equation (\ref{e9}). Therefore, evaluating this equation for each node in $LS^{q+c}$ we can realize the likelihood of a location being the real location of the queried set $LS^{q+c}$. Finally, we borrow the concept of entropy to understand the uncertainty in the data calculated as
\begin{align}
h= -\sum_{y=1}^{k^{q+c}} \textrm{Pr}(l^{q+c}_y&=r^{q+c}|LS^q,...,LS^{q+c-1})\nonumber \\ &\log_{2}\textrm{Pr}(l^{q+c}_y=r^{q+c}|LS^q,...,LS^{q+c-1}).
\end{align}
We call $h$, the transition-entropy of the set $LS^{q+c}$ with respect to the previous $c$ queried location sets $LS^q,...,LS^{q+c-1}$. As it will be demonstrated in simulation results, the proposed transition-entropy metric will indicate the susceptibility of the locations in $LS^{q+c}$ to be identified as dummies or real location of the user based on previous the queried location sets of the trajectory. The algorithm to calculate the transition-entropy metric is formally presented in Algo. \ref{Algo3}.

\begin{figure}[h]
\centering
\includegraphics[scale=0.41]{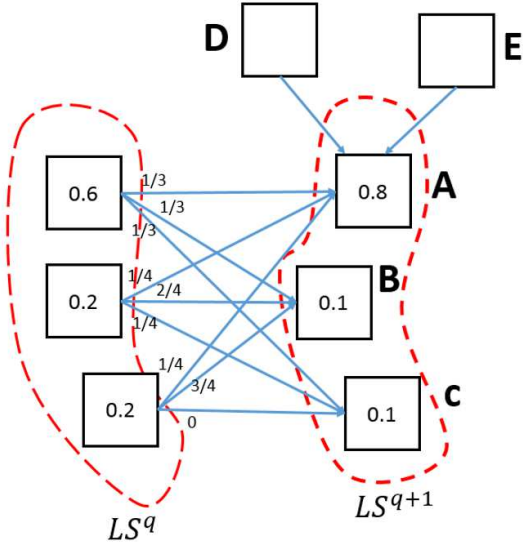}
\hspace{1em}
\centering
\caption{An example of two consecutive queried location sets.}
\label{f4}
\end{figure}

Calculation of transition-entropy is only based on the query probability of initial location set $LS^{q}$ and the transition entropies throughout the trajectory. It is important to understand why the query probability of the other locations on the trajectory are not considered in the calculation of the transition-entropy metric. This can best be understood by an example. Fig. \ref{f4} demonstrates a user requesting a LBS in two consecutive queries. The numbers written on the nodes indicate the normalized query probability of the locations and the numbers written on the edges indicate the normalized probability of that transition. Assume we want to calculate the transition-entropy metric for $LS^{q+1}$ based on the previous queried location set $LS^{q}$. The purpose of the example is to illustrate why the posterior probabilities calculated by the previous queries for $LS^{q+1}$ is more reliable than the query probability of the locations in $LS^{q+1}$. First, let us calculate the posterior probabilities of $LS^{q+1}$ and its entropy. The posterior probabilities according to the equation (\ref{e12}) can be written as

\begin{align}
\textrm{Posterior probability of A being the true location}=\\
\dfrac{3}{5}\times \dfrac{1}{3}+\dfrac{1}{5}\times \dfrac{1}{4}+\dfrac{1}{5}\times \dfrac{1}{4}=\dfrac{6}{20}\\
\textrm{Posterior probability of B being the true location}=\\
\dfrac{3}{5}\times \dfrac{1}{3}+\dfrac{1}{5}\times \dfrac{2}{4}+\dfrac{1}{5}\times \dfrac{3}{4}=\dfrac{9}{20}\\
\textrm{Posterior probability of C being the true location}=\\
\dfrac{3}{5}\times \dfrac{1}{3}+\dfrac{1}{5}\times \dfrac{1}{4}=\dfrac{5}{20}
\end{align}
According to the query probabilities of $LS^{q+1}$ the location $A$ is more likely to be the real location as it has a significantly higher query probability, but looking at the posterior probabilities calculated for the location set we can see that based on $LS^{q}$, location $B$ is more probable to be the real location of the user. This discrepancy can be explained by looking at what the actual meaning of query probability is. The query probability indicates the number of times a location has been called but does not specify if it is been called after any particular location. Therefore, although location $A$ has been called more times than the other locations in $LS^{q+1}$, most of these queries perhaps have been made consecutively after locations $E$ and $D$ which are not a member of the location set $LS^{q}$. Hence, it can be seen that the posterior probabilities are more credible as they are considering the number of times queries made after prior location set $LS^{q}$.

\section{Viterbi Attack}
\label{section: v}
The Viterbi algorithm is a well-known dynamic programming algorithm proposed in $1967$ by the authors in \cite{viterbi}. Initially, it was used for convolutional codes, but then it found numerous applications such as exploring the most likely sequence of hidden states in Hidden Markov Models (HMMs). For a given graph, the aim of the algorithm is to find the shortest path or so-called the most likely path. The most likely path is usually referred to as the Viterbi path. The Viterbi algorithm provides several features which distinguishes this algorithm from others existing algorithms for this purpose. The most important characteristic of the algorithm can be mentioned as low computational complexity. Here, we design an attack based on the Viterbi algorithm and name it Viterbi attack since the principal idea behind the attack is inspired by the Viterbi algorithm. The proposed Viterbi attack can significantly endanger the location privacy of the users if it is not considered in the design of the dummy generation algorithms. The adversary can exploit the accessed side information such as transition probabilities to compromise the location privacy of the users by conducting Viterbi attack. As it will be demonstrated in simulations, for a user traveling an even short trajectory the Viterbi attack can successfully identify many of the real location. In the following, We adopt and explain how the Viterbi algorithm can be considered as a threat to the location privacy of users.

Given the queried location sets $LS^q$, $LS^{q+1}$,...,$LS^{q+c}$ corresponding to a trajectory of length $c+1$ of a user, an attacker seeks to find the most probable state sequence to compromise the location privacy of the user. Here $l^{j}_i \in LS^{j}$ is referred to as a state of the location set $LS^{j}$. The desired state sequence of the adversary would be $(r^q,r^{q+1},...,r^{q+c})$ where for $j= q\textrm{ to } q+c$, $r^{j}$ refers to the true location of the queried set $LS^{j}$. We define $\mu(m+1,u)$ to be the maximum probability of a state sequence with the length of $m+1$ given as $z^q,z^{q+1},...,z^{q+m}$ where $z^{j} \in LS^{j}$ and $z^{q+m}=u\in LS^{q+m}$. This function can be expressed mathematically as
\begin{equation}\label{e19}
\mu(m+1,u) = \underset{z^{q:q+m}|z^{q+m}=u}{\mathlarger{max}}\textrm{Pr}(z^{q+m}=r^{q+m}),
\end{equation}
where for each $u\in LS^q$ the initial value of the $\mu$ function is set to
\begin{align}\label{e20}
\mu(0,u) = \textrm{Pr}(u=r^q),
\end{align}
in which as the most credible information for the first queried location set is the query probability, $\textrm{Pr}(u=r^q)$ is calculated via equation (\ref{e10}). Starting from the second queried location set the most probable path can be calculated recursively as
\begin{equation}\label{e21}
\mu(m+1,u) = \underset{u'\in LS^{q+m-1}}{\mathlarger{max}}\mu(m,u')\textrm{Pr}(u'\rightarrow u).
\end{equation}

\begin{algorithm}[t]
\DontPrintSemicolon 
\textbf{Input:} queried location sets $LS^q$, $LS^{q+1}$,...,$LS^{q+c}$ and the normalized query probability for the location set $LS^q$\\
\textbf{Initialization:} .\\
\For {$1\leq u \leq k^q$} {
$\mu(0,u)= \textrm{Pr}(l_u^q=r^q)$\\
$pointer(0,u)=0$
}

\For {$1\leq j \leq c$} {
\For {$1\leq u \leq k^{q+j}$} {
$\mu(j,u) = \underset{u'\in LS^{q+j-1}}{\mathlarger{max}}\mu(j-1,u')\textrm{Pr}(u'\rightarrow u)$\\
$pointer(j,u)\leftarrow \textrm{state of }\underset{u'\in LS^{q+j-1}}{\mathlarger{max}}\mu(j-1,u')$
}
}
$EstState[c]= \textrm{state of } max(\mu(c,:))$\\
\For {$c-1\geq j \geq 0$} {
$EstState[j]=pointer(j+1,EstState[j+1])$
}
\textbf{Output:} $EstState$.\\
\caption{The algorithm of the proposed Viterbi attack.}
\label{Algo5}
\end{algorithm}

The formal presentation of Viterbi attack is given in the Algo. \ref{Algo5}. The algorithm starts by setting the initial values of the $\mu$ array to their normalized query probability in lines $3-6$. An array called $pointer$ is used to keep track of the most likely state of the previous queried location set as the most probable path is calculated in lines $7-12$. Finally, the most probable path is chosen and the corresponding states are returned as output.

\section{The proposed algorithms to improve location privacy of users}
\label{section:AlgoS}
In this section, we start by proposing two algorithms for improving the transition-entropy metric. The algorithms are independent of the method used for generating the dummies. For the purpose of explanation, the underlying dummy generation algorithm is set to DLS in our work. The first proposed algorithm is based on exhaustively searching for the desired dummy set and the second algorithm follows a greedy approach for selection of the dummies. We continue by proposing an algorithm called robust dummy generation (RDG) which can significantly increase the privacy of the users against the Viterbi attack while maintaining the high performance in terms of transition-entropy and cell-entropy.

\subsection{Exhaustive Search Algorithm}

\begin{algorithm}[t]
\DontPrintSemicolon 
\textbf{Input:} $k^{q+1}$, the location set $LS^q=\{l^q_1,l^q_2,...,l^q_{k^q}\}$ in $q$-th query, and the location set $LS^{q+1}=\{l^{q+1}_1\}$ which only includes the real location of the user at $(q+1)$-th query. \\
\textbf{Output:} the location set $LS^{q+1}$ which includes the real location and $(k^{q+1}-1)$ dummies.\\
\textbf{Initialization:} .\\
$D\leftarrow$ generate a pool of $4k^{q+1}$ dummies using the DLS algorithm\\

$\{S_1,S_2,...S_m\}\leftarrow$ choose $m$ distinct $(k^{q+1}-1)$-subsets of $D$ \\
\For {$1\leq y \leq m$} {
$S_y\leftarrow S_y\cup \{l^{q+1}_1\}$\\
$h_y\leftarrow$ calculate transition-entropy of $S_y$\\
$H\leftarrow H\cup \{h_y\}$\\
}
\For {$1\leq y \leq m$} {
{\If {$h_y$ is the maximum number in $H$ } {\textbf{return} $S_y$\\ exit;}
}
}

\caption{The proposed exhaustive search algorithm for location privacy preservation of the users.}
\label{Algo3}
\end{algorithm}

Suppose that at time $t^q$ the user has made its $q$-th query for the location set of $LS^q=\{l^q_1,l^q_2,...,l^q_{k^q}\}$ which includes the real location and its associated dummies. As the user changes its location and makes his $(q+1)$-th query at time $t^{q+1}$, assuming $k^{q+1}$-anonymity for the user, we wish to generate the location set $LS^{q+1}=\{l^{q+1}_1,l^{q+1}_2,...,l^{q+1}_{k^{q+1}}\}$ to maximize the transition-entropy metric. The idea is to generate a pool of dummies instead of only $k^{q+1}-1$ fake locations which have similar cell-entropy to the real location of the user and choosing $k^{q+1}-1$ subsets of the dummy location pool for evaluation of their transition-entropy performance. The formal description of the proposed method for generating the dummies of the set $LS^{q+1}$ is explained in the Algorithm \ref{Algo3}. The procedure starts by generating a pool of $4k^{q+1}$ dummies using the DLS algorithm and assigning them to set $D$. Then, $m$ distinct subsets of $D$ are chosen each with $(k^{q+1}-1)$ members, which will form a complete $k^{q+1}$ set of locations by addition of the real location ($l^{q+1}_1$). Finally, the transition-entropy of each set is calculated with respect to $LS^{q}$, and the set with the maximum transition-entropy is returned as the $(q+1)$-th query set.

The proposed exhaustive search algorithm considers the extra side information incorporated in this paper. As it will be demonstrated in simulation results, the algorithm provides a significantly better transition-entropy performance compared to the existing algorithms while maintaining the traditional cell-entropy metric near optimal.

\subsection{Greedy Algorithm}

\begin{algorithm}[t]
\DontPrintSemicolon 
\textbf{Input:} $k^{q+1}$, the location set $LS^q=\{l^q_1,l^q_2,...,l^q_{k^q}\}$ in $q$-th query, and the location set $LS^{q+1}=\{l^{q+1}_1\}$ which only includes the real location of the user at $(q+1)$-th query. \\
\textbf{Output:} the location set $LS^{q+1}$ which includes the real location and $(k^{q+1}-1)$ dummies.\\
\textbf{Initialization:} .\\
$D\leftarrow$ generate a pool of $4k^{q+1}$ dummies using the DLS algorithm\\

\For {$1\leq member \leq k^{q+1}-1$} {
$entropy=zeros(1\times |D|)$\\
\For {$1\leq d \leq |D|$} {
$LS^{q+1}=LS^{q+1}\cup \{D_d\}$\\
$entropy[d]\leftarrow \textrm{transition entropy of }LS^{q+1} \textrm{ w.r.t. } LS^q$\\
$LS^{q+1}=LS^{q+1}- \{D_d\}$\\
}
$NewMember \leftarrow \{\textrm{member of } D \textrm{ which maximize } entropy\}$
$LS^{q+1}=LS^{q+1}\cup \{NewMember\}$\\
$D=D- \{NewMember\}$\\
}
\textbf{return} $LS^{q+1}$

\caption{The proposed greedy algorithm for location privacy preservation of the users.}
\label{Algo4}
\end{algorithm}

Although the Exhaustive search algorithm can significantly improve the location privacy of the users, the high computational overhead is a major drawback of the algorithm. The computational cost of the exhaustive search algorithm is in the order of $O(mk^2)$ where if no bound is selected for $m$, its value is $C^{4k}_k$. It can be seen that the implementation of such algorithm can be time-consuming. Therefore, in order to decrease the computation complexity, we propose a greedy approach which can achieve an order of $O(k^3)$.

Following the same setup explained in exhaustive search algorithm we aim to generate the set $LS^{q+1}$ in a way to maximize the transition-entropy with respect to the previous location set $LS^q$. The principal idea behind the greedy algorithm is to choose the members which maximize the transition-entropy one by one and add them to the $LS^{q+1}$ instead of looking at all the possible combinations and the transition-entropy they achieve. The algorithm starts by generating a pool of dummies using the DLS algorithm. The DLS algorithm has been chosen due to its robust performance in terms of cell-entropy, the algorithm is applicable for other dummy generation methods as well. Initially, the location set $LS^{q+1}$ only includes the real location of the user at $(q+1)$-th query. The next member is added by trying out all the members in $D$ and calculating the transition-entropy of $LS^{q+1}$ including that member and choosing the one which maximizes the transition-entropy. Then, we move to the third member and the same procedure is repeated until all the $k^{q+1}-1$ dummies are chosen. The greedy algorithm is formally presented in Algorithm \ref{Algo4}.

\subsection{RDG Algorithm}

In this subsection, we propose RDG algorithm in which the aim is to increase the resilience against the Viterbi attack while maintaining the cell-entropy and transition-entropy as high as the currently existing algorithms.

The algorithm is based on the idea of posterior probabilities introduced as part of the derivation of the transition-entropy. We explain the algorithm for generation of $LS^{q+1}$ from the queried location set $LS^{q}$. If $LS^{q}$ is the initial query of the user from the LBS provider, then, the initial posterior probabilities are set to the normalized query probability of the locations in $LS^{q}$; otherwise, the posterior probabilities are calculated from equation (\ref{ee13}). In the algorithm, posterior probabilities are assigned to an array called $weight$.

The algorithm starts by the generation of a pool of dummies using the DLS algorithms based on the real location of $LS^{q+1}$. Using DLS algorithm to generate the pool of dummies will ensure the high performance of the algorithm in terms of the cell-entropy. From our experiments, setting the pool size to four times of the $k^{q+1}$ would still keep the cell-entropy quite high while resulting a robust performance in terms of the transition-entropy and Viterbi attack resilience. Next, the algorithm continues by employing a greedy approach to add the most suitable dummies for the location set $LS^{q+1}$. For choosing the $i$-th member of the set $LS^{q+1}$, each of the remaining dummies in the pool is checked one by one. A criterion chosen here is based on maximizing the entropy for the array $weight$. For each member $u \in LS^{q+1}$, the $weight$ array is calculated as
\begin{equation}\label{e22}
weight(2,u) = \underset{u'\in LS^{q}}{\mathlarger{max}}weight(1,u')\textrm{Pr}(u'\rightarrow u).
\end{equation}
The weight array is chosen to be a two-dimensional array to distinguish between the weights for different location sets. For each member of the dummy pool, its weight is calculated followed by the entropy of the weight array. After calculation of the entropy for all the possible members, the member which results in maximum entropy is chosen as a next member of $LS^{q+1}$. The process continues until all the $k^{q+1}-1$ dummies of $LS^{q+1}$ are chosen. Note that before calculation of the entropy the weights are normalized to make the accumulation of the probabilities add up to one. The algorithm has been designed to provide a high cell-entropy and transition-entropy privacy for the users while protecting them from the Viterbi attack on trajectories.

\begin{algorithm}[t]
\DontPrintSemicolon 
\textbf{Input:} $k^{q+1}$, the location set $LS^q=\{l^q_1,l^q_2,...,l^q_{k^q}\}$ in $q$-th query, and the location set $LS^{q+1}=\{l^{q+1}_1\}$ which only includes the real location of the user at $(q+1)$-th query. \\
\textbf{Output:} the location set $LS^{q+1}$ which includes the real location and $(k^{q+1}-1)$ dummies.\\
\textbf{Initialization:} .\\

\For {$1\leq u \leq k^q$} {
$weight(1,u)\leftarrow \textrm{Posterior probability of }l_u^q$\\
}

$D\leftarrow$ generate a pool of $4k^{q+1}$ dummies using the DLS algorithm\\
\For {$1\leq member \leq k^{q+1}-1$} {
$entropy=zeros(1\times |D|)$\\
\For {$1\leq d \leq |D|$} {
$LS^{q+1}=LS^{q+1}\cup \{D_d\}$\\
\For {$1\leq u \leq k^{q+1}$} {
$weight(2,u) = \underset{u'\in LS^{q}}{\mathlarger{max}}weight(1,u')\textrm{Pr}(u'\rightarrow u)$\\
}
normalize $weight(2,:)$\\
$entropy[d]\leftarrow \textrm{entropy of }weight(2,:)$\\
$LS^{q+1}=LS^{q+1}- \{D_d\}$\\
}
$NewMember \leftarrow \{\textrm{member of } D \textrm{ which maximize } entropy\}$
$LS^{q+1}=LS^{q+1}\cup \{NewMember\}$\\
$D=D- \{NewMember\}$\\
}

\textbf{return} $LS^{q+1}$

\caption{The proposed greedy algorithm for location privacy preservation of the users.}
\label{Algo6}
\end{algorithm}

\section{Performance Evaluation}
\label{section:Performance}

\subsection{Experiment Setup}

In our experiment, we use the data collected by Geolife project \cite{d1,d2,d3}, which includes the GPS trajectories of $182$ users from April 2007 to August 2012 in Beijing, China. The dataset contains the GPS logs of the users including $17,621$ trajectories with a total distance of $1,292,951 km$. There are two main advantages distinguishing Geolife dataset for our work. Firstly, the recorded data aside from monitoring the daily routines of the users, such as going to work or home, includes trajectories involving the sports activities like hiking and cycling. Secondly, many of the recorded trajectories are tagged with a transportation mode, which indicates the use of various means of traveling from bus and car to airplane and train.

We have conducted our experiments on $1km\times 1km$ central part of the Beijing map with the resolution of $0.01km\times 0.01km$ for each grid cell. The location privacy requirement ($k$) of the users are investigated for the values $2$ to $30$. For each value of $k$, the algorithms are repeated $3000$ time to ensure the reliability of the results. Although the proposed algorithm and metric can be used for the users who have varying location privacy requirements in consecutive queries of the LBS, for the sake of comparison, we have assumed that the $k$ value stays the same in consecutive calls for the LBS. Additionally, the experiments are performed on a PC with a $3.40$GHz core-i7 Intel processor, $64$-bit Windows $7$ operating system, and $8.00$GB of RAM. Moreover, Python program is used to implement the algorithms.

\begin{figure}[t]
\centering
\includegraphics[scale=.88,height=6cm]{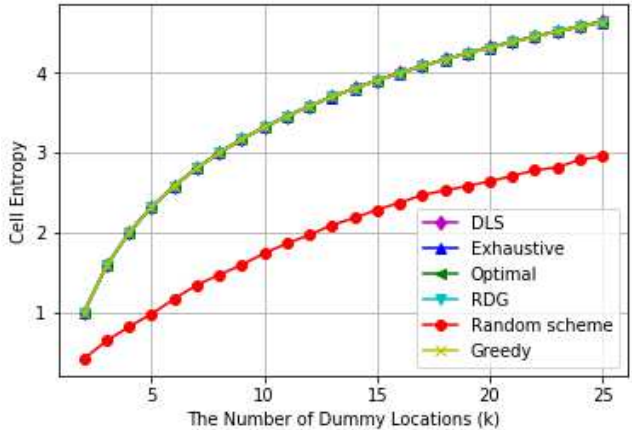}
\hspace{1em}
\centering
\caption{Comparison of algorithms in terms of cell-entropy for different values of $k$.}
\label{f5}
\end{figure}

\subsection{Performance Analysis}
In this section, we evaluate the performance of the proposed algorithms and metrics through an extensive number of experiments. The desired outcome of the experiments is to show that the proposed RDG algorithm can withhold the currently established metric of cell-entropy \cite{entropy} while increasing the performance in terms of the proposed metric transition-entropy and providing the users with a high privacy preservation against the developed Viterbi attack. We start by analyzing the performance of the algorithms in terms of cell-entropy, followed by transition-entropy analysis and investigating their performance against Viterbi attack.

\subsubsection{Cell-entropy performance evaluation}

In order to calculate the cell-entropy metric, the adversary records the number of times each cell has been queried over time, and using this information calculates the query probability of each cell. Once the dataset including the real location and dummies are submitted to the server, the adversary can calculate the cell-entropy of the user. A higher value for the cell-entropy indicates more uncertainty in finding the real location or recognizing the dummies. Therefore, maximum cell-entropy is desirable to maintain the $k$-anonymity of the users.

Fig. \ref{f5} represents the comparison of different algorithms in terms of cell-entropy. The optimal value is achieved when the $k$ locations queried form the LBS provider all have the same probability of $\dfrac{1}{k}$, or equivalently, the location set has the cell-entropy of $h= \log_{2}k$. The optimal value is the target for all the algorithms since it is the maximum entropy that a location set can achieve. In the random scheme \cite{baseline}, the dummies are generated randomly which expectedly results in a lower cell-entropy compared to the other algorithms. As it can be seen in the figure, the DLS algorithm achieves near-optimal performance in terms of the cell-entropy. Therefore, the adversary is unable to compromise the $k$-anonymity of the user from the stationary set of locations submitted to the server using the available query probabilities. The exhaustive, greedy and RDG algorithms can also achieve near-optimal performance which indicates that in all the algorithms the adversary is unable to identify the dummy location by exploiting the cell-entropy. It must be noted that proposed algorithms here is adaptable to any dummy generation algorithm, therefore, the reason for a high cell-entropy performance of the proposed algorithms is that we have chosen DLS as our base. Hence, if other algorithms are chosen, the cell-entropy performance must be evaluated for them as well to ensure the robust performance in terms of the cell-entropy.

\begin{figure}[t]
\centering
\includegraphics[scale=.88,height=6.05cm]{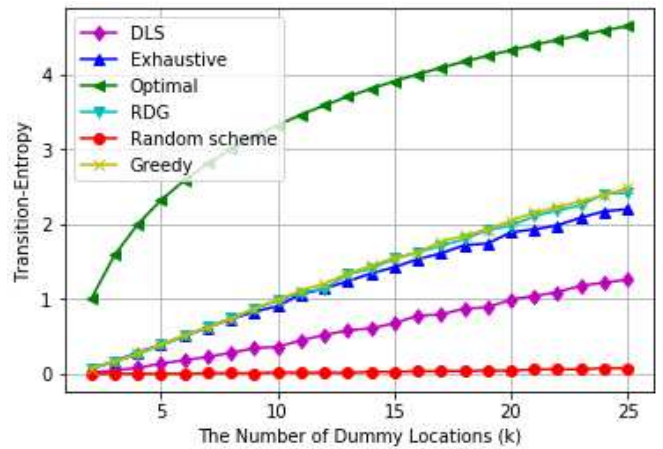}
\hspace{1em}
\centering
\caption{Comparison of algorithms in terms of transition-entropy for different values of $k$ in trajectory of length $2$.}
\label{f6}
\end{figure}

\subsubsection{Transition-entropy performance evaluation}

The currently established cell-entropy metric only considers the location privacy for the stationary set of queried locations submitted to the LBS server, but overlooks the fact that the adversary has access to the trajectories traveled by the users as well. The adversary can use the likelihood of traveling different paths between the consecutive location sets, and infer with a high probability that many of the submitted locations are dummies, which leads to failure in preserving location privacy requirements of the users. Fig. \ref{f6} compares the performance of different algorithms in terms of the transition-entropy for a path length of two. For all the algorithms, based on the value of $k$, two consecutive location sets are generated, each including the real location and its associated dummies. To make the experiments as realistic as possible, the real location movements are chosen randomly from the recorded trajectories in the dataset.

\begin{figure}[t]
\centering
\includegraphics[scale=.88,height=6.05cm]{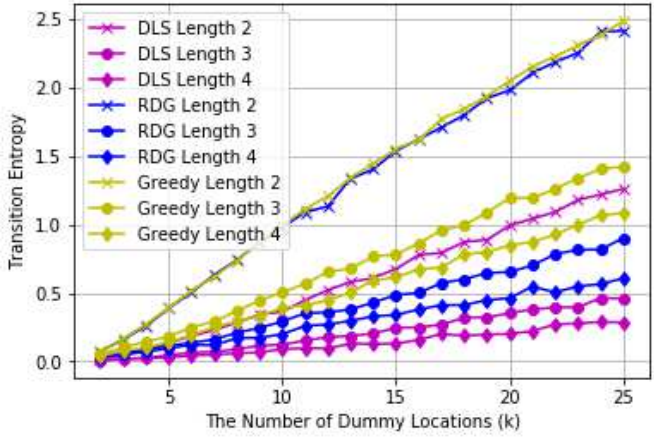}
\hspace{1em}
\centering
\caption{Comparison of algorithms in terms of transition-entropy for different values of $k$ in trajectories of length $2,3$ and $4$.}
\label{f7}
\end{figure}

\begin{figure}

\begin{subfigure}
\centering
\includegraphics[scale=.91,height=6.2cm]{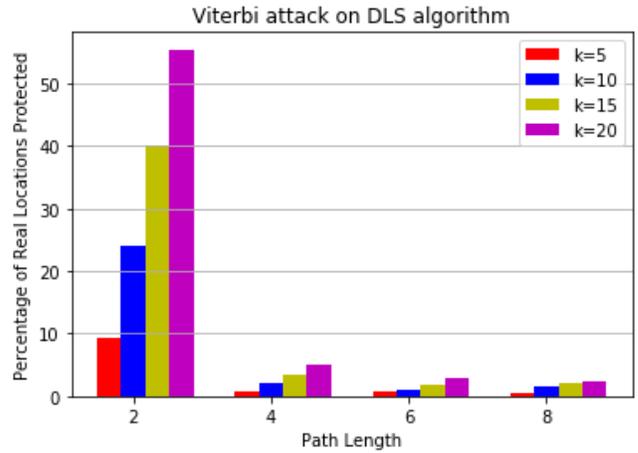}
\hspace{1em}
\centering
\end{subfigure}

\begin{subfigure}
\centering
\includegraphics[scale=.91,height=6.2cm]{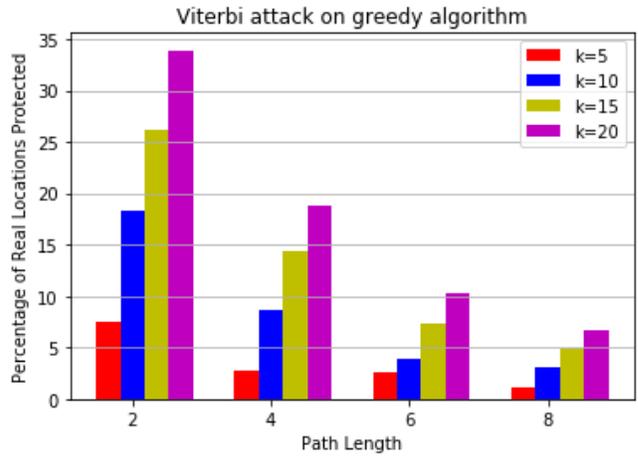}
\hspace{1em}
\centering
\end{subfigure}

\begin{subfigure}
\centering
\includegraphics[scale=.91,height=6.2cm]{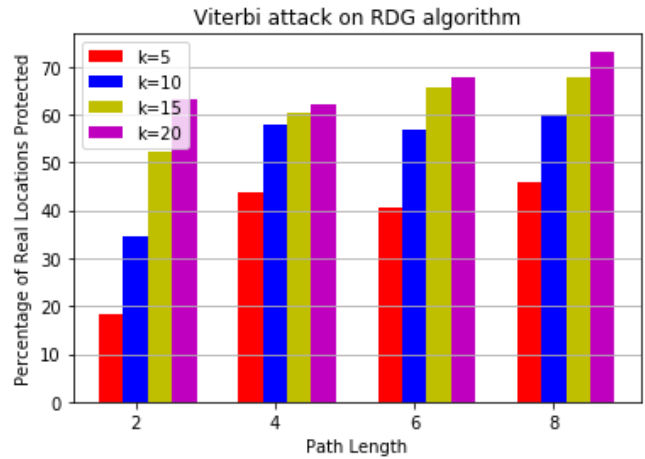}
\hspace{1em}
\centering
\end{subfigure}
\caption{The performance evaluation of DLS, greedy, and RDG algorithms against Viterbi attack considering various path lengths and privacy requirement $k$}
\label{f8}
\end{figure}

The optimal value in Fig. \ref{f6} corresponds to a scenario in which all the members of the second location set are equally likely to be called consecutively after the members of the first location set. The optimal values can be calculated in a similar way as the optimal number for the cell-entropy for different values of $k$. As it can be seen from the figure, the random scheme has a very poor performance which means that the adversary can easily recognize most of the dummies from the transition-entropy even for the two consecutive location sets queried by the user. The first point to notice in the figure is that although DLS algorithm achieved a near-optimal performance in terms of cell-entropy, transition-entropy performance indicates that the adversary can compromise the location privacy of the users by calculating the posterior probabilities. The transition-entropy of the proposed algorithms in this paper can be seen to significantly improve the transition-entropy performance, almost improving the performance more than twice as high as the DLS algorithm. In other words, the likelihood of compromising the $k$-anonymity requirement is decreased by the proposed algorithms which leads to a higher location privacy for the users of LBSs. The exhaustive algorithm can be seen to achieve a little worse performance compared to the RDG and greedy. This lower performance is due to setting an upper bound for the number of sets chosen for calculation of the transition-entropy instead of going through them all which will become highly computational when the pool size of dummies is large. The performance of RDG can be seen to significantly high compared to the other algorithms.

Fig. \ref{f7}, extends our analysis of transition-entropy for trajectories with higher length. The crucial inference from the graph is that as more number of locations are queried from the LBS provider, the transition-entropy reduces. This simulation result corresponds to the theoretical analysis that by having more information the adversary is able to calculate the posterior probabilities more accurately which results in less uncertainty for the adversary to identify the real location of the users. The previous algorithm DLS can be seen to have a very low transition-entropy compared to the proposed algorithms greedy and DLS. Therefore, our proposed algorithms are viable in increasing the transition-entropy of the users while maintaining the cell-entropy to near-optimal performance. It must be noted that the greedy and RDG are able to increase the transition-entropy for different dummy generation algorithms without depending on what the underlying algorithm for the generation of the pool of dummies is. Therefore, a better algorithm than DLS algorithm can cause the performance to improve as the greedy and RDG algorithms increased the transition-entropy of DLS algorithm.

\subsection{Performance of Algorithms Against Viterbi Attack}

In this subsection, we analyze the performance of our proposed algorithms against the designed Viterbi attack. The performance analysis is given in Fig. \ref{f8}. Considering the extensive side information we incorporated in this paper, the Viterbi attack would be a significantly threatening privacy issue for the users of LBSs. Looking at the percentage of real locations protected in the Viterbi attack on DLS algorithm, it can be seen that, for instance, in a trajectory of length $8$ the adversary is able to identify almost all the real locations of the users. This shows that although in a single request of LBS from the server the locations are protected using existing dummy generation algorithms, in trajectories the side information that the adversary has, can cause the compromised LBS provider to almost identify all the real locations. The apparent trend for all the path lengths is that increasing the number of dummies can improve the preservation of location privacy, but this increase is not sufficient even for trajectories of length two.

The second algorithm considered in Fig \ref{f8} is the greedy algorithm proposed in our work to increase the transition-entropy of the dummy generation algorithms. Although the algorithm prevents the inference of real locations based on transition-entropy, it is not capable of providing location privacy against the Viterbi attack conducted by the adversary. The performance of greedy algorithm against Viterbi attack gets worse as more number of queries are made from the LBS as the adversary will have more accurate information from the history of data. On the other hand, looking at the performance analysis of RDG, it can be seen that as the algorithm tends to confuse the adversary more and more in each requested query from the LBS provider, having larger trajectories the difference between the real path and estimated path of the Viterbi attack becomes larger. RDG algorithm is able to protect at least $50$ percent of the user queried locations if the $k$-anonymity criterion is set to $15$ or larger.

\section{Conclusions}
\label{section: Conclusions}
In this work, we incorporated new side information which can be exploited by the adversary to compromise the location privacy of the users. We proposed a metric called transition-entropy to evaluate the performance of the dummy-based algorithms and quantified the currently existing metric cell-entropy. The metric is based on the transitions between the locations in the map and considers the deplorable effect of new side information on location privacy of the users. To improve the transition-entropy metric two general approaches were proposed to increase the transition-entropy for a given dummy generation algorithm. Furthermore, we developed an attack model based on the Viterbi algorithm on location privacy of the users, followed by proposing an algorithm called RDG to increase the performance in terms of the cell-entropy and transition-entropy while protecting the users against Viterbi attack. Finally, numerous experiments were performed on real-world data to analyze the performance of the algorithms.

\bibliographystyle{IEEEtran}
\bibliography{paper_ref}

\end{document}